\documentclass{PoS}
\usepackage{sidecap}
\usepackage{tikz}
\usepackage{floatrow}
\usepackage{microtype}
\usepackage[utf8]{inputenc} 

\def\Xint#1{\mathchoice
   {\XXint\displaystyle\textstyle{#1}}%
   {\XXint\textstyle\scriptstyle{#1}}%
   {\XXint\scriptstyle\scriptscriptstyle{#1}}%
   {\XXint\scriptscriptstyle\scriptscriptstyle{#1}}%
   \!\int}
   
\def\XXint#1#2#3{{\setbox0=\hbox{$#1{#2#3}{\int}$}
     \vcenter{\hbox{$#2#3$}}\kern-.5\wd0}}

\def\dashint{\Xint-}

\title{$\pi\pi$ scattering in a renormalized Hamiltonian
  matrix~\thanks{This work is supported by the Spanish MINECO and
    European FEDER funds (grant FIS2017-85053-C2-1-P) and Junta de
    Andaluc\'{\i}a (grant FQM-225). M.G.R has been supported in part
    by the Spanish MINECO's Juan de la Cierva-Incorporaci\'on programme,
    Grant Agreement No. IJCI-2017-31531.  }}

\ShortTitle{$\pi\pi$ scattering in a renormalized Hamiltonian matrix}

\author{\speaker{Mar\'ia G\'omez-Rocha}\\
        Universidad de Granada and Instituto Carlos I de F\'isica Te\'orica y Computacional\\
        E-mail: \email{mgomezrocha@ugr.es}}

\author{Enrique Ruiz Arriola\\
        Universidad de Granada and Instituto Carlos I de F\'isica Te\'orica y Computacional\\
        E-mail: \email{earriola@ugr.es}}

\abstract{A Wilsonian approach to $\pi\pi$ scattering based in the
  Glazek-Wilson Similarity Renormalization Group (SRG) for Hamiltonians
  is analyzed in momentum space up
  to a maximal CM energy of $\sqrt{s}=1.4$ GeV. To this end, we
  identify the corresponding relativistic Hamiltonian by means of the
  3D reduction of the Bethe-Salpeter equation in the Kadyshevsky
  scheme, introduce a momentum grid and provide an isospectral
  definition of the phase-shift based on a spectral shift of a
  Chebyshev angle. We also propose a new method to integrate the SRG
  equations based on the Crank-Nicolson algorithm with a single step
  finite difference so that isospectrality is preserved at any step of
  the calculations.  We discuss issues on the unnatural high momentum
  tails present in the fitted interactions and reaching far beyond the
  maximal CM energy of $\sqrt{s}=1.4$ GeV and how these tails can be
  integrated out explicitly by using Block-Diagonal generators of the
  SRG.}

\FullConference{Light Cone 2019 - QCD on the light cone: from hadrons to heavy ions - LC2019\\
		16-20 September 2019\\
		Ecole Polytechnique, Palaiseau, France}

\begin{document}

\section{Introduction}

Scattering experiments provide an important source of information
regarding hadronic interactions. This is so, for instance, for
$\pi\pi$ scattering where accurate determinations of phase-shifts
exist~\cite{GarciaMartin:2011cn} but give no clue on how the $3\pi$
system possibly explaining $\omega$ and $A_1$ should be
handled. Hamiltonian methods involving potentials become very valuable
when treating the mentioned hadronic interactions as a few-body
problem where one pursues the determination of bound states and
resonances of multi-hadron systems in terms of the corresponding
potentials. However, existing model $\pi\pi$ potentials that fit the
interaction display an annoying long momentum tail
(cf.~\cite{Mathelitsch:1986ez,Gomez-Rocha:2019rpj}), which involves
high energy scales not natural to the physical problem.  In this work
we address this issue using the similarity renormalization group
(SGR) approach to Hamiltonian
dynamics~\cite{Glazek:1993rc,Glazek:1994qc,Wilson:1994fk} and employ a
new method for calculating phase shifts that has been
investigated recently~\cite{Gomez-Rocha:2019xum,Gomez-Rocha:2019rpj}.

\section{The Kadyshevsky Hamiltonian: Phase-shifts and the spectral-shift method}
\label{sec:Kad}

The theoretical determination of scattering phase-shifts from potentials requires solving an integral equation, which can be tackled
numerically applying a discretization of integrals using a momentum
grid. We choose the Kadyshevsky
equation~\cite{Kadyshevsky:1967rs}, which allows for a Hamiltonian
formulation
and can be easily extended to the more general case of the three-body
problem~\footnote{The standard relativistic approach to describe
  $\pi\pi$ scattering is the Bethe-Salpeter equation (BSE). In
  practice, due to the complications that the four-dimensional nature
  of the integrals presents, one employs judicious 3D reductions of
  the BSE, which are closer in spirit to the nonrelativistic
  Lippmann-Schwinger equation.}. The reaction matrix in the
Kadyshevsky version is related to the scattering phase shifts through:
\begin{eqnarray}
R_l (p',p,\sqrt{s}) &=& V_l(p',p) 
\ + \ 
\dashint_0^\infty dq \, \frac{q^2}{4 E_q^2} 
\frac{V_l(p',q) R_l (q,p, \sqrt{s})}{\sqrt{s}-2 E_q }  \ ,\quad
-\tan \delta_l(p) = \frac{\pi}{8} \frac{p}{E_p} R_l (p,p,\sqrt{s} ) 	\ . \nonumber \\ 
\label{eq:Kad-PV}
\end{eqnarray}
The Hamiltonian corresponding to the Kadyshevsky equation
Eq.~(\ref{eq:Kad-PV}) is given by
\begin{eqnarray}
  H \Psi_l(p)  \equiv 2 E_p \Psi(p) + \int_0^\infty dq \frac{q^2}{4 E_q^2} v_l(p,q)
  \Psi_l (q)  \ ,
  \label{eq:Kad-Hamiltonian}
\end{eqnarray}
which in general needs to be solved numerically. We choose the Gauss-Chebyshev quadrature: 
\begin{eqnarray}
p_n &=& \frac{\Lambda_{\rm num}}{2}\left[ 1- \cos \phi_n \right] \ , \quad 
w_n \ = \   \frac{\Lambda_{\rm num}}{2} \frac{\pi}{N} \phi_n \  ,
\quad \phi_n={\pi\over N}\left(n - {1\over 2}\right)
\label{eq:pnwn}
\end{eqnarray}
where $n=1, \dots, N$, and $\phi_n$ is the {\it Chebyshev angle}. On
the momentum grid, Eq.~(\ref{eq:Kad-Hamiltonian}) becomes 
\begin{eqnarray}
2 E_n \Psi_n 
  + \sum_k w_k \frac{p_k^2}{4 E_k^2} V_{n,k} \Psi_k = \sqrt{s} \Psi_n  \ .
\label{eq:KadH}  
\end{eqnarray}
Phase shifts can be determined from the spectrum of the Hamiltonian,
as it has been explained by DeWitt~\cite{DeWitt:1956be} and Fukuda and
Newton~\cite{Fukuda:1956zz}, who linked the shift produced in the
spectrum after introducing the interaction with the scattering phase
shifts appearing in the $S$-matrix. The relation is based on the use
of an equidistant energy and momentum grids, respectively.  For our
Gauss-Chebyshev grid, which is equidistant in the Chebyshev angle, the
corresponding relation becomes~\cite{Gomez-Rocha:2019xum,Gomez-Rocha:2019rpj}
\begin{eqnarray}
  \delta_n= - \pi \frac{\Phi_n- \phi_n}{\Delta \phi_n} \equiv - \pi \frac{\Delta \Phi_n}{\Delta \phi_n} \ ,
\label{eq:phishift}  
\end{eqnarray}
where,
$\Delta \phi_n={\pi\over N}$, and the ``distorted'' angles $\Phi_n$ are
calculated inverting in Eqs.~(\ref{eq:pnwn}) and replacing $p_n$ by $P_n$,
which is extracted from the Hamiltonian eigenvalue
$\sqrt{s}=2E_n=\sqrt{m^2+P_n^2}$.

\begin{figure*}[h]

\floatbox[{\capbeside\thisfloatsetup{capbesideposition={right,center},capbesidewidth=6.5cm}}]{figure}[\FBwidth]
{\caption{$\pi\pi$ scattering phase shifts in the $S0$ calculated using the $\phi$-shift method using a grid of $N=$50 points (blue dots) compared with the model fit, (green, smooth line) and experimental data taken from~\cite{GarciaMartin:2011cn} (red dots with error bars). See all channels in~\cite{Gomez-Rocha:2019rpj}.}\label{fig:exp}}
{\includegraphics[scale=0.55]{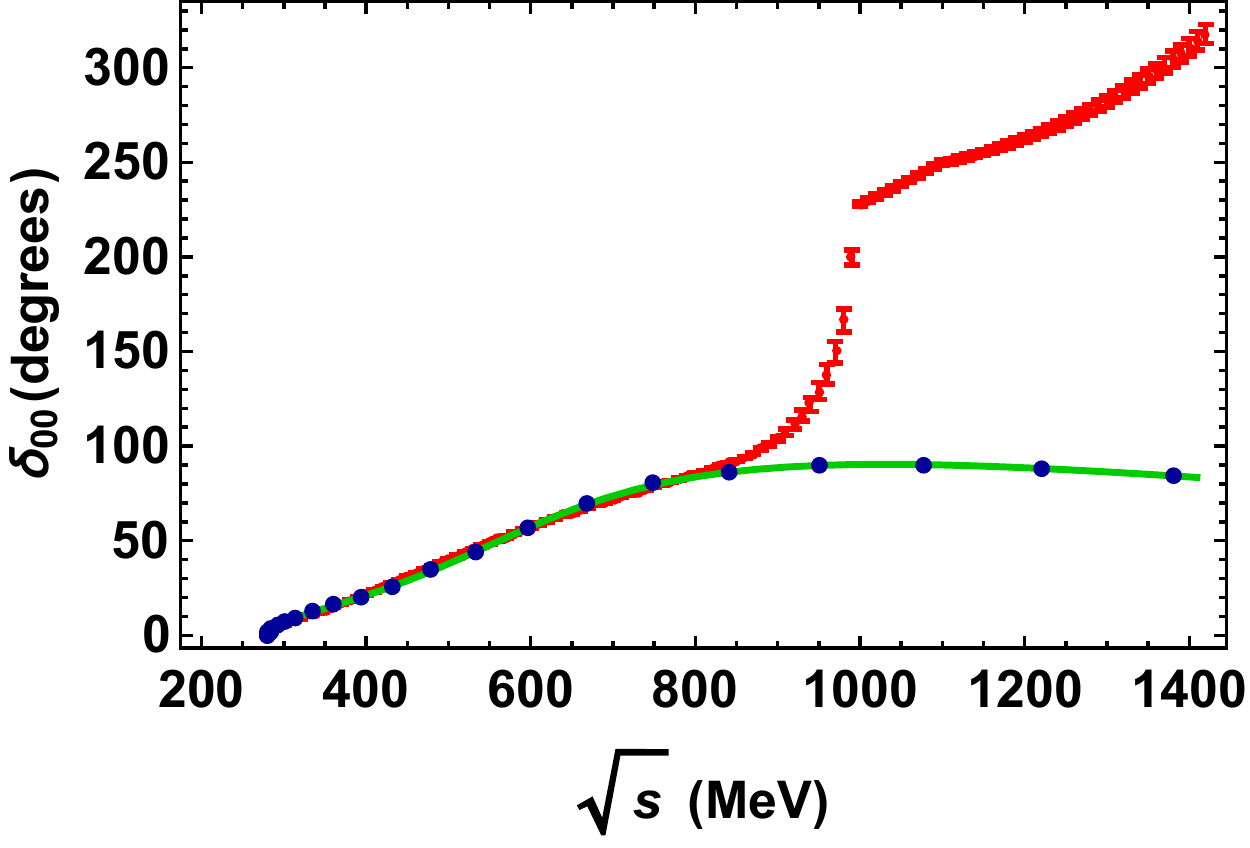}}
\end{figure*}

\begin{figure*}[ht]
\centering 
\includegraphics[scale=0.55]{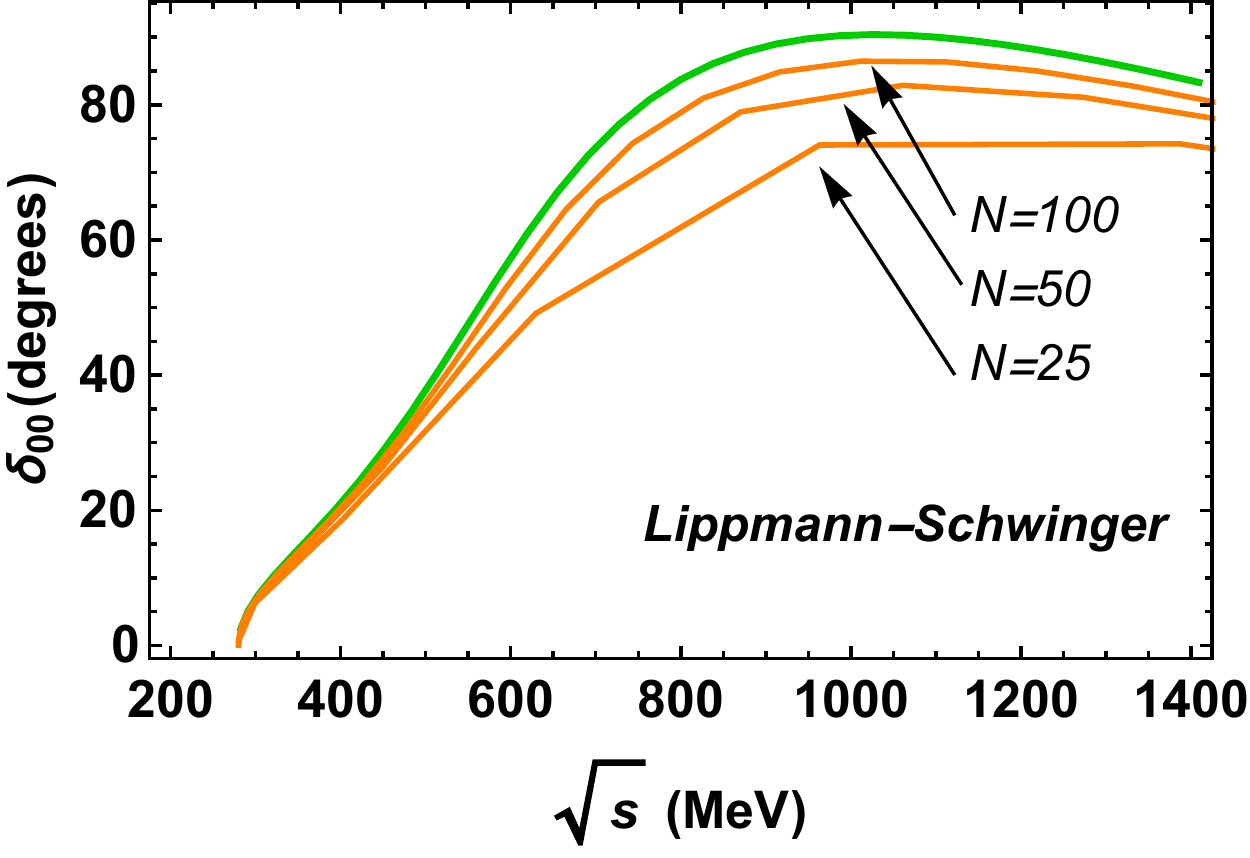}
\includegraphics[scale=0.55]{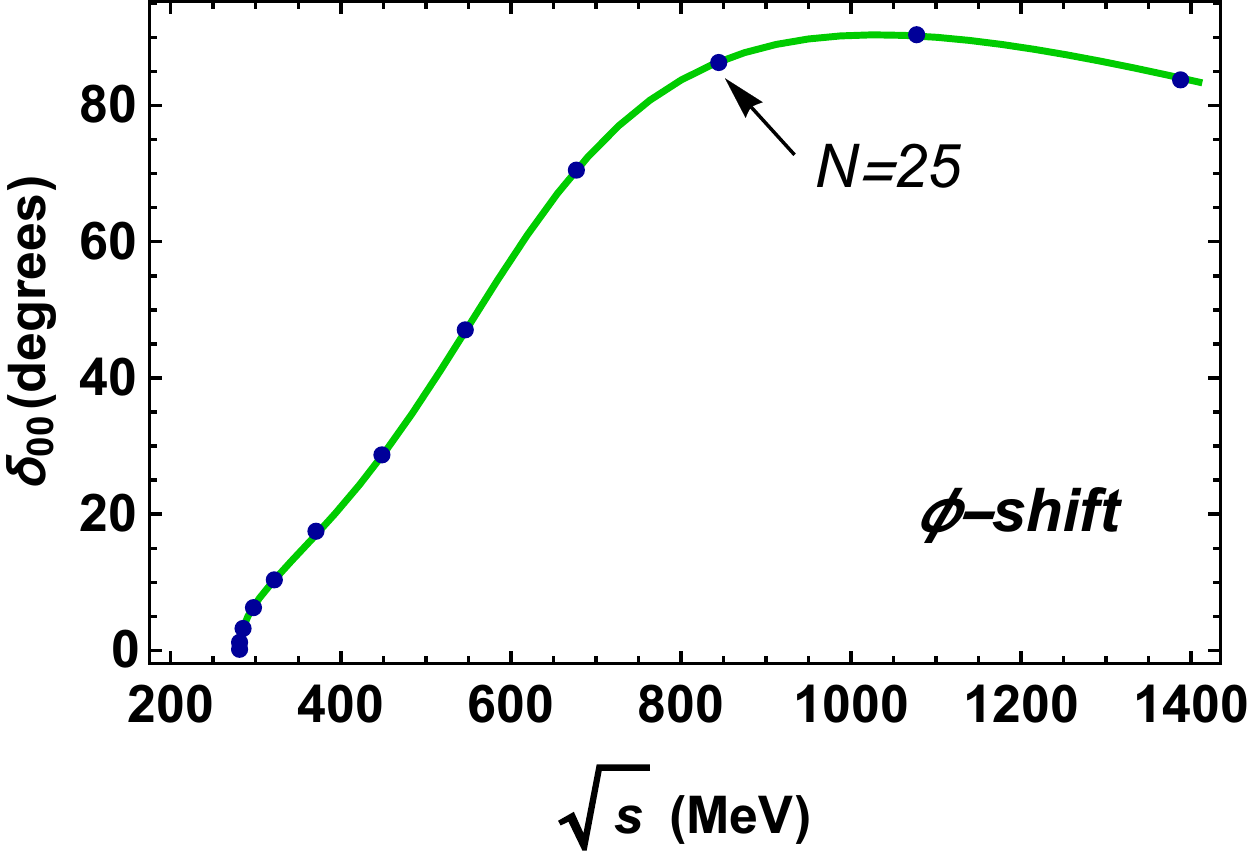}
\caption{Phase shifts for the $S0$ channel calculated using the LS equation  versus the $\phi$-shift method. Left panel: results obtained using LS equation for three grids of different number of points. Right panel: $\phi$-shift method for a grid of $N=$25 points. The green, smooth line is the model fit in both graphics.  See a deeper analysis with all channels in~\cite{Gomez-Rocha:2019rpj}.} 
\label{Fig:LS-vs-phi-shift}
\end{figure*}

For our discussion we take the $\pi\pi$ potentials introduced in
Ref.~\cite{Mathelitsch:1986ez}. In Figure~\ref{fig:exp} we show our
$\phi$-shift results compared with the model
fit~\cite{Mathelitsch:1986ez,Gomez-Rocha:2019rpj} and with the
experimental
data~\cite{GarciaMartin:2011cn}~\footnote{Phenomenologically, the
  steep raise of the $\pi\pi$ phase-shift in the $S0$ channel is due to the onset of the
  subthreshold $\bar K K$ channel not included here for
  simplicity.}. As it can be seen, the prediction is very
accurate. Even for a grid with a relative small number of points, the
results are much better and less sensitive to discretization effects
than the results obtained from standard Lippann-Schwinger (LS)
equation (cf. Figure~\ref{Fig:LS-vs-phi-shift} and~\cite{Gomez-Rocha:2019rpj} for details). Note also the
improvement with respect to the \textit{momentum} or
\textit{energy-shift} method described
in~\cite{Gomez-Rocha:2019zkz,Gomez-Rocha:2019rpj}.

%

\section{Similarity renormalization group and $\pi\pi$ scattering}
\label{sec:srg}

\begin{figure*}[ht]
\centering 
\floatbox[{\capbeside\thisfloatsetup{capbesideposition={right,center},capbesidewidth=6.5cm}}]{figure}[\FBwidth]
{\caption{Diagonal matrix elements of the block-diagonal evolved
     potential for the S0 channel compared to the original one. It can be done analogously for all channels~\cite{Gomez-Rocha:2019zkz,Gomez-Rocha:InPreparation}}\label{Fig:comparisons}}
{\includegraphics[scale=0.65]{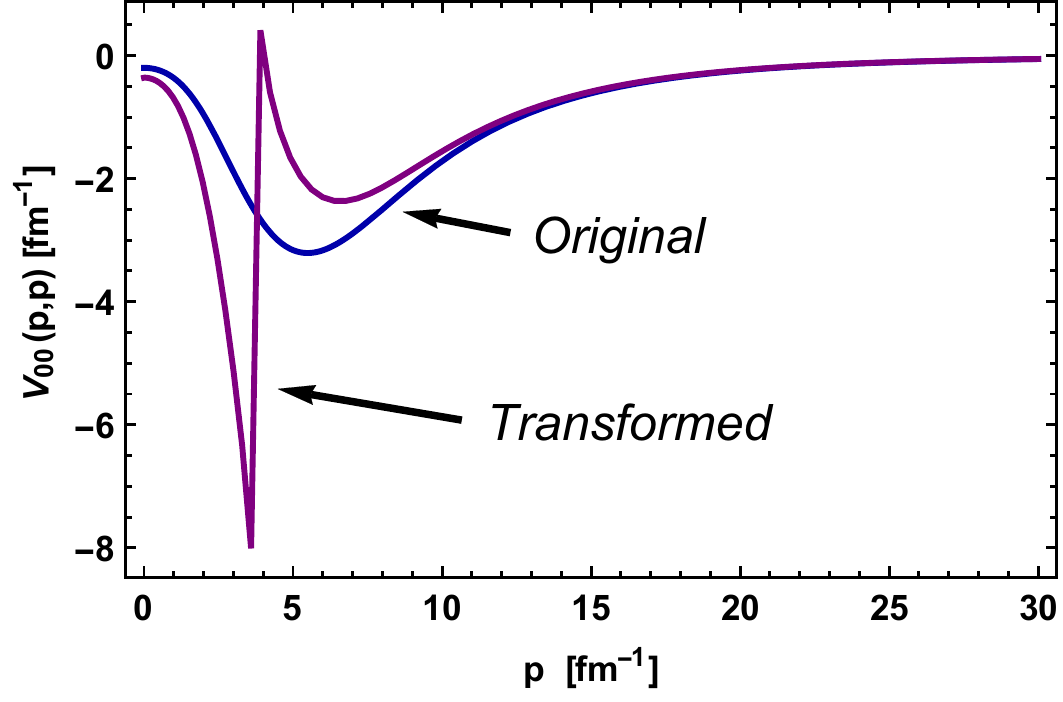}}
\end{figure*}

\begin{figure*}[h]
  \includegraphics[scale=0.45]{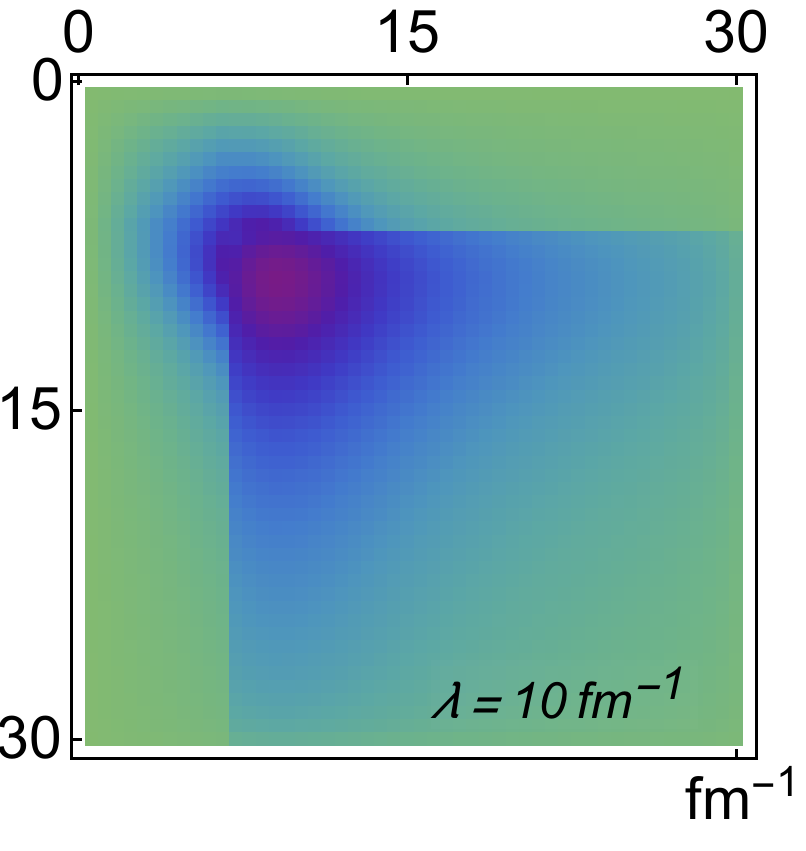}
  \includegraphics[scale=0.45]{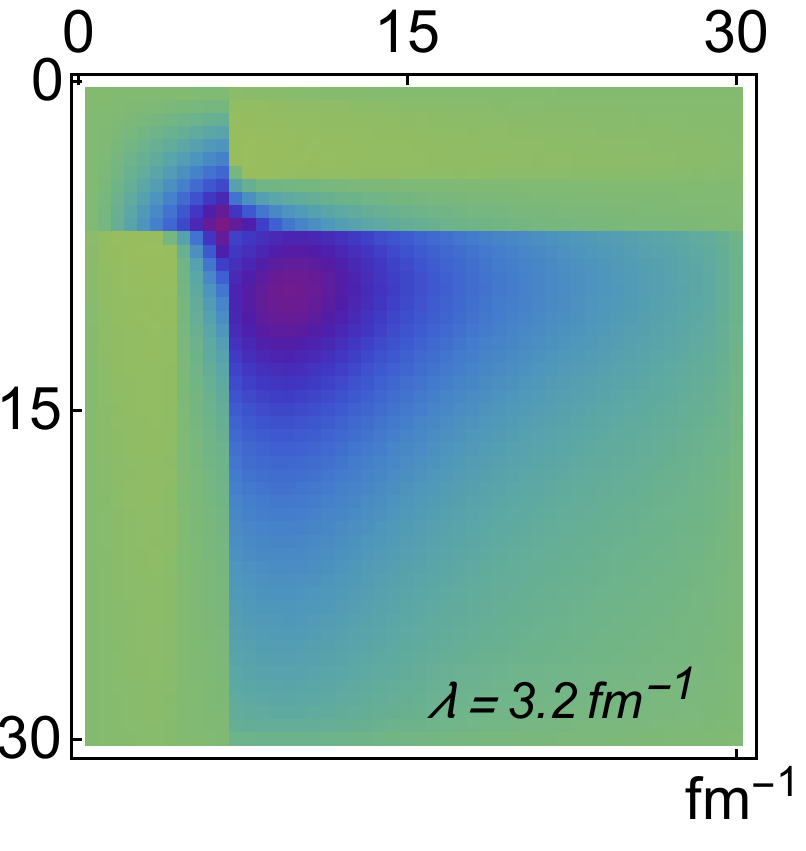}
  \includegraphics[scale=0.45]{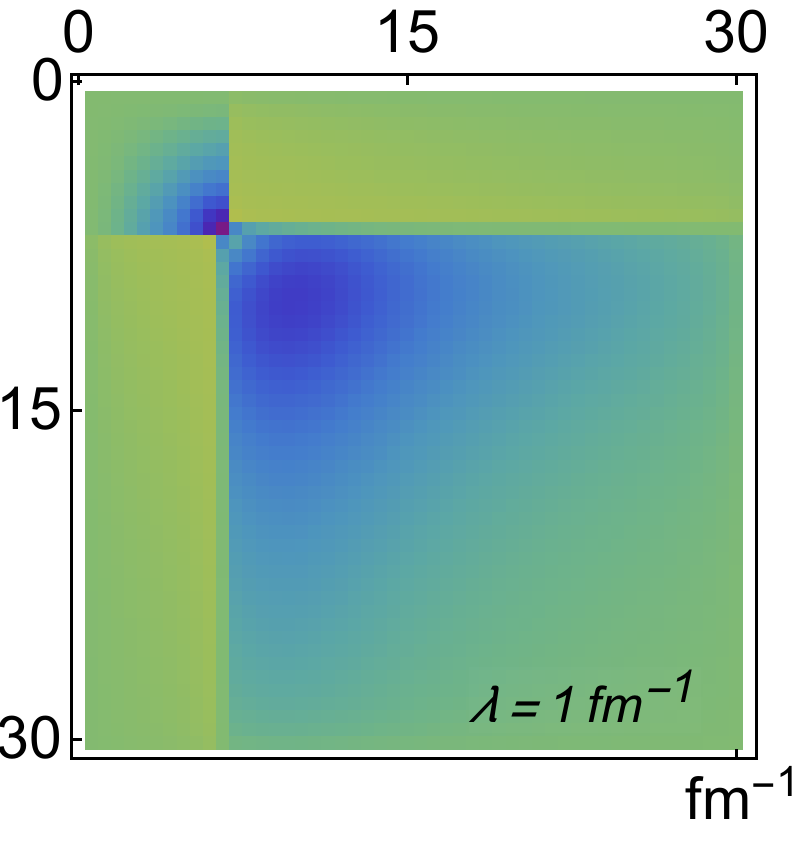}
  \includegraphics[scale=0.45]{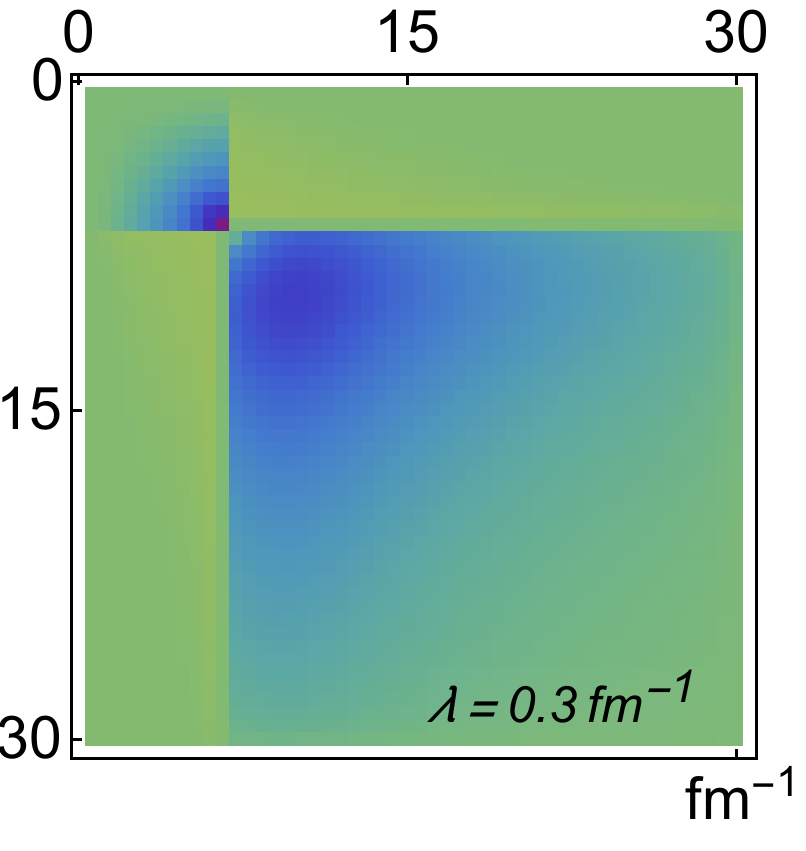}
  \caption{Block-diagonal evolved Hamiltonian in the $S0$ channel corresponding to different values of $\lambda$.
  The calculation is made using a grid of $N=$50 points.}
\label{Fig:potentialevolved}
\end{figure*}

The phenomenological potentials present very long tails (up to 30 GeV,
see Fig.~\ref{Fig:comparisons}), an annoying fact that appears
unnatural considering the experimental region reaches only $\sim 1.5$
GeV.
To handle this we invoke the
SRG~\cite{Glazek:1993rc,Glazek:1994qc,Wilson:1994fk} which has been
used mostly in nuclear physics to tame the nuclear core, and resting on the possibility to apply a (scale dependent)
unitary transformation on the initial Hamiltonian, evolving it in a
continuous way into a more convenient basis {\it preserving} the
eigenvalues, and hence the phase-shifts.
The SRG scale-dependence is dictated by the equation 
\begin{eqnarray}
\frac{d H_t}{dt}= [[G_t,H_t],H_t] \ , 
\label{eq:srg}
\end{eqnarray}
where $t$ is the renormalization-group parameter and $G_s$ is the
generator which determines the basis into which the {\it running}
Hamiltonian, $H_t$, is transformed. One customary choice is to use a
diagonal-like Wilson generator, $G_t=T$, which transforms the
Hamiltonian matrix into a narrow band-diagonal one using in practice
the Crank-Nicolson method (see e.g.~\cite{Gomez-Rocha:2019zkz}). Here we consider instead a block-diagonal generator $G=PHP+QHQ$, by means
of the orthogonal projectors, $P=\theta(\Lambda-p)$ and $Q=
\theta(p-\Lambda)$, which define two subspaces separated by a cut-off
$\Lambda$ and which transforms the matrix into a block-diagonal
one. Figure~\ref{Fig:potentialevolved} illustrates the evolution of
the Hamiltonian matrix corresponding to the $S0$ channel potential of
Section~\ref{sec:Kad} up to different values of $\lambda=1/\sqrt{t}$.
The block-diagonal generator decouples two subspaces, below and beyond
the cutoff $\Lambda\sim 1400$ MeV, as it is visible in the lower-right
panel of Figure~\ref{Fig:potentialevolved}. The matrix elements
relevant to the experimental region are now contained in the small
block, and the eigenvalues of such a submatrix are the ones of the
complete matrix with values below the cutoff {\it only}, thus
eliminating higher energies explicitly.

\section{Conclusions and outlook}
\label{sec:conclusions}
We have used the $S0$ channel in $\pi\pi$ scattering to illustrate how
the SRG can be employed to treat the problem of long tails potential
in hadronic physics. Using a block-diagonal generator, we have
constructed an effective Hamiltonian that allows to decouple two
energy regions and select the matrix block relevant to the
experimental region. We have presented a
method~\cite{Gomez-Rocha:2019xum,Gomez-Rocha:2019rpj} for calculating
phase shifts based on the spectrum modification that appears after
introducing the interaction, which is obtained given a momentum grid.

Once the effective Hamiltonian has been constructed and the different
energy regions have been decoupled, only the small matrix is needed to
study the physical problem, and a corresponding potential in the
reduced region can be identified, thus explicitly ignoring any
ultraviolet effects.  From the evolved matrix and using
Eq.~(\ref{eq:KadH}) in inverse way, we can identify the corresponding
effective potential, whose diagonal is plotted in
Figure~\ref{Fig:comparisons}. Since the reduced subspace of the
effective potential (i.e. from 0 to about 5 fm$^{-1}$) yields the same
phase shifts as the whole original Hamiltonian, we can simply
consider the reduced space, which contains much less points, to reproduce experiments and to treat further physical
problems that requires such a potential. An example is the three-body
problem, e.g., the $\omega$ and $A_1$ resonances decaying to three
pions, that includes the two-body potential as a component in its
formulation. Moreover, the fact that the $\phi$-shift method
reproduces quite accurately the experimental results for rather coarse
momentum grids, can advantageously be used as an alternative tool that
reduces the computational cost in a situation where the binding in the
three-body system is dominated by an infrared cut-off corresponding
to the finite size of the three-body system.

\bibliographystyle{h-elsevier}
\bibliography{newrefs,srg-rel,refs}

\end{document}